\documentclass[%
 reprint,
superscriptaddress,
amsmath,amssymb,
]{revtex4-2}
\usepackage{graphicx}
\usepackage{dcolumn}
\usepackage{bm}
\usepackage{lineno}
\usepackage{hyperref}
\hypersetup{
            colorlinks=true,
            linkcolor=blue,
            anchorcolor=blue,
            citecolor=blue
            }

\begin{document}

\title{ Optimization of thermo-spin voltage in vertical
nanostructures by geometrical means }

\author{Fupeng Gao}
\affiliation{School of Microelectronics $\&$ State Key Laboratory for Mechanical
Behavior of Materials, Xi'an Jiaotong University, Xi'an 710049, China }

\author{Shaojie Hu}
\email[]{hu.shaojie@phys.kyushu-u.ac.jp} 
\affiliation{Department of Physics, Kyushu University, 744 Motooka, Fukuoka, 819-0395, Japan}

\author{Dawei Wang}
\affiliation{School of Microelectronics $\&$ State Key Laboratory for Mechanical Behavior of Materials, Xi'an Jiaotong University, Xi'an 710049, China }

\author{Takashi Kimura}
\email[]{t-kimu@phys.kyushu-u.ac.jp}
\affiliation{Department of Physics, Kyushu University, 744 Motooka, Fukuoka, 819-0395, Japan}

\date{\today} 

\begin{abstract}
The thermo-spin conversion provides new concepts for further developing the green energy-harvesting technology because spin can be controlled with minimal energy in nanostructures. Through theoretical analysis of thermo-spin generation, transportation and conversion in ferromagnet/non-ferromagnet/heavy metal (FM/NM/HM) vertical structures, we found that the output transverse thermo-spin voltage is independent of the structure's width, but varies in a linear function with the structure's length.  To validate our predictions, we fabricated the thermo-spin devices with a CoFeAl/Cu/Pt structure.  Our results indicate that FM/NM/HM structures can be utilized to design flexible thermo-spin conversion devices.
\end{abstract}

\maketitle
\section{Introduction}

The thermoelectric effect is a promising technology for efficiently utilizing heat energy to achieve a future green and low-carbon society \cite{2008_Bell,DiSalvo1999,Sundarraj2014}. The discovery of thermally excited spin current has promoted research on the thermoelectric conversion of heat to electricity via spin current, leading to the development of a new branch of spintronics called spin caloritronics\cite{2010Bauer}. This emerging field provides new concepts for further developing green energy-harvesting technology because spin can be controlled with minimal energy in nanostructures \cite{Hu2014,2019Mizuguchi}. The most representative spin-thermoelectric effects are the longitudinal spin Seebeck effect (LSSE) in ferromagnetic insulator/Heavy metal (FI/HM) structures\cite{uchida2010observation,2016Wu_H,Meier_2015}, spin-dependent Seebeck effect (SDSE) in ferromagnet/non-ferromagnet (FM/NF) structures \cite{Slachter2010ThermallyDS,Hu2014,choi2015thermal} and the anomalous Nernst effect (ANE) in ferromagnet \cite{hu2013anomalous,2015Sakuraba,HuangAPL2020,2021Journal}. The longitudinal structure has been used in most recent studies of the transverse thermo-electric effect because the hybrid structure is suitable for thermoelectric applications due to its simple configuration.  Figure 1(a) shows thermo-spin current, which is generated by LSSE or SDSE, can generate a transverse electric field by inverse spin Hall effect (ISHE) in a heavy metal (HM) layer \cite{kimura2007room,2006Saitoh,2006Valenzuzela,2008takeshi,takahashi2008spin}. The electric field could be expressed as: $\textbf{E}_x^{\rm HM}= \theta_{\rm SH}^{\rm HM}(\textbf{m}_y\times \textbf{j}_{\rm sz}^{\rm HM})\rho_{\rm HM}$. $\theta_{\rm SH}^{\rm HM}$ is the spin Hall angle in HM. $\textbf{j}_{\rm sz}^{\rm HM}$ is the thermo-spin current along z direction in HM. $\rho_{\rm HM}$ is the resistivity of HM.
One of the unique characteristics of the transverse thermo-spin device is the denominator of the figure of merit, thermal conductivity k and resistivity $\rho$, which are free from the Wiedemann–Franz law. This means that it can be optimized for each layer separately \cite{Uchida_2014,Uchida2016ThermoelectricGB}.
The anomalous Nernst effect also generates a transverse potential in a ferromagnetic (FM), as shown in Fig.1(b). The ANE can be regarded as the thermal effect of the anomalous Hall effect in ferromagnetic metals \cite{RevModPhys.82.1539}. 
The ANE, which is analyzed through the spin-charge conversion model, can be phenomenologically explained by considering the interaction of SDSE-induced thermo-spin currents with the inverse spin Hall effect in the ferromagnetic\cite{fang2016scaling}.
So, the electric field in ferromagnet could be expressed as: $\textbf{E}_x^{FM}= \theta_{\rm SH}^{\rm FM}(\textbf{m}_y\times \textbf{j}_{\rm sz}^{\rm FM})\rho_{\rm FM}$. $\theta_{\rm SH}^{\rm FM}$ is the spin Hall angle in FM. $\textbf{j}_{\rm sz}^{\rm FM}$ is the thermo-spin current along z direction in FM. $\rho_{FM}$ is the resistivity of FM.
In addition, ANE could be enhanced or inhibited by adjusting the composition structure of ferromagnets \cite{ramos2015unconventional,fang2016scaling,2017Kannan}.

\begin{figure}[htp]
    \centering 
    \includegraphics[width=3.2in]{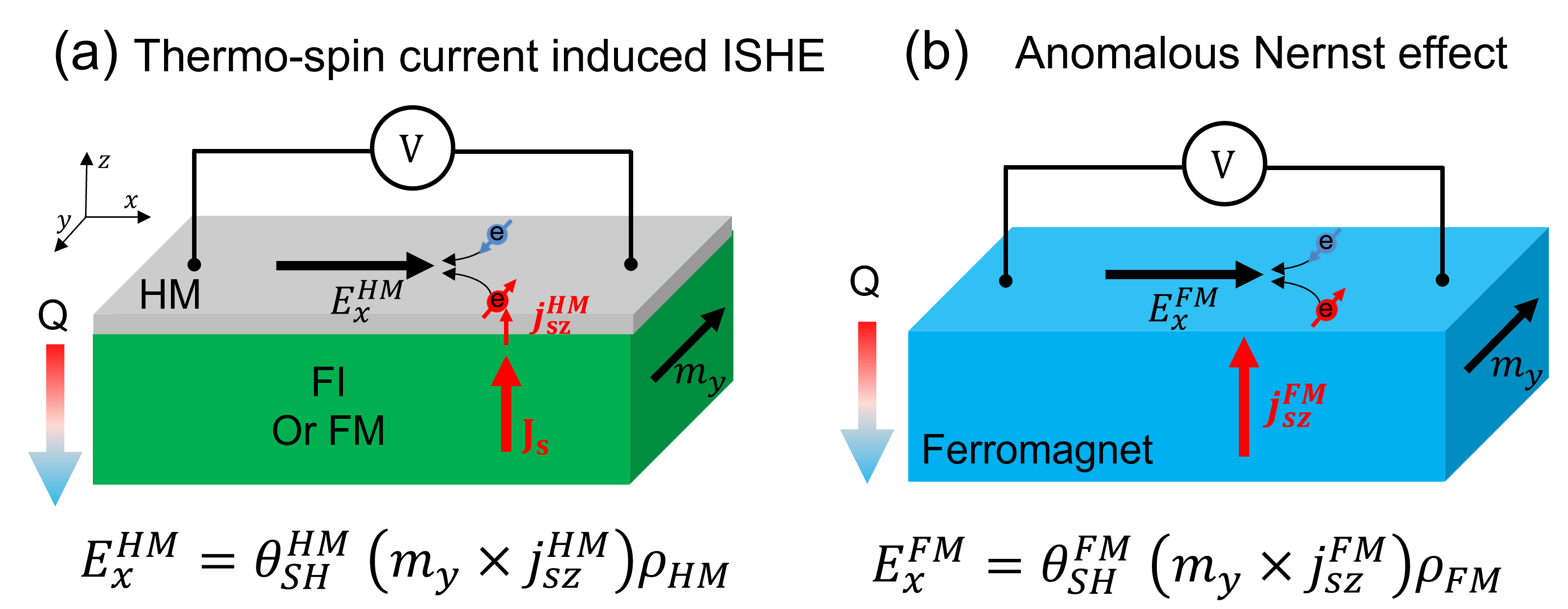}
    \caption{(a) Schematic diagram of inverse spin Hall effect due to the LSSE or SDSE induced thermo-spin current in HM. Q is the heat current. (b) Schematic diagram of the anomalous Nernst effect in a ferromagnet by considering the interaction of thermo-spin currents with the inverse spin Hall effect.}
    \label{1}
\end{figure}
The transverse thermoelectric property is one of the most notable features of such effects, where a longitudinal temperature gradient generates a transverse electric field. This unique advantage provides much greater flexibility in device design compared to conventional thermoelectric effects. While studies of the longitudinal spin Seebeck effect have mainly focused on ferromagnetic insulators or ferrimagnets, optimizing internal resistance in such systems remains challenging. Ferromagnetic metals offer more options for realizing maximum power output through the flexible design of internal resistance. The generation efficiency of thermo-spin current and inverse spin Hall angle are also crucial parameters for achieving substantial output power and voltage.
The CoFeAl(CFA)/Cu hybrid nanostructure has been confirmed to have high efficiency for thermo-spin injection properties \cite{Hu2014,hu2014significant,nomura2017efficient}. In this manuscript, we designed a thermo-spin electric device with an efficient thermal spin injection hybrid structure CoFeAl/Pt by inserting Cu layer to study thermal spin injection and conversion properties.
\begin{figure*}[htp]
    \centering 
    \includegraphics[width=6in]{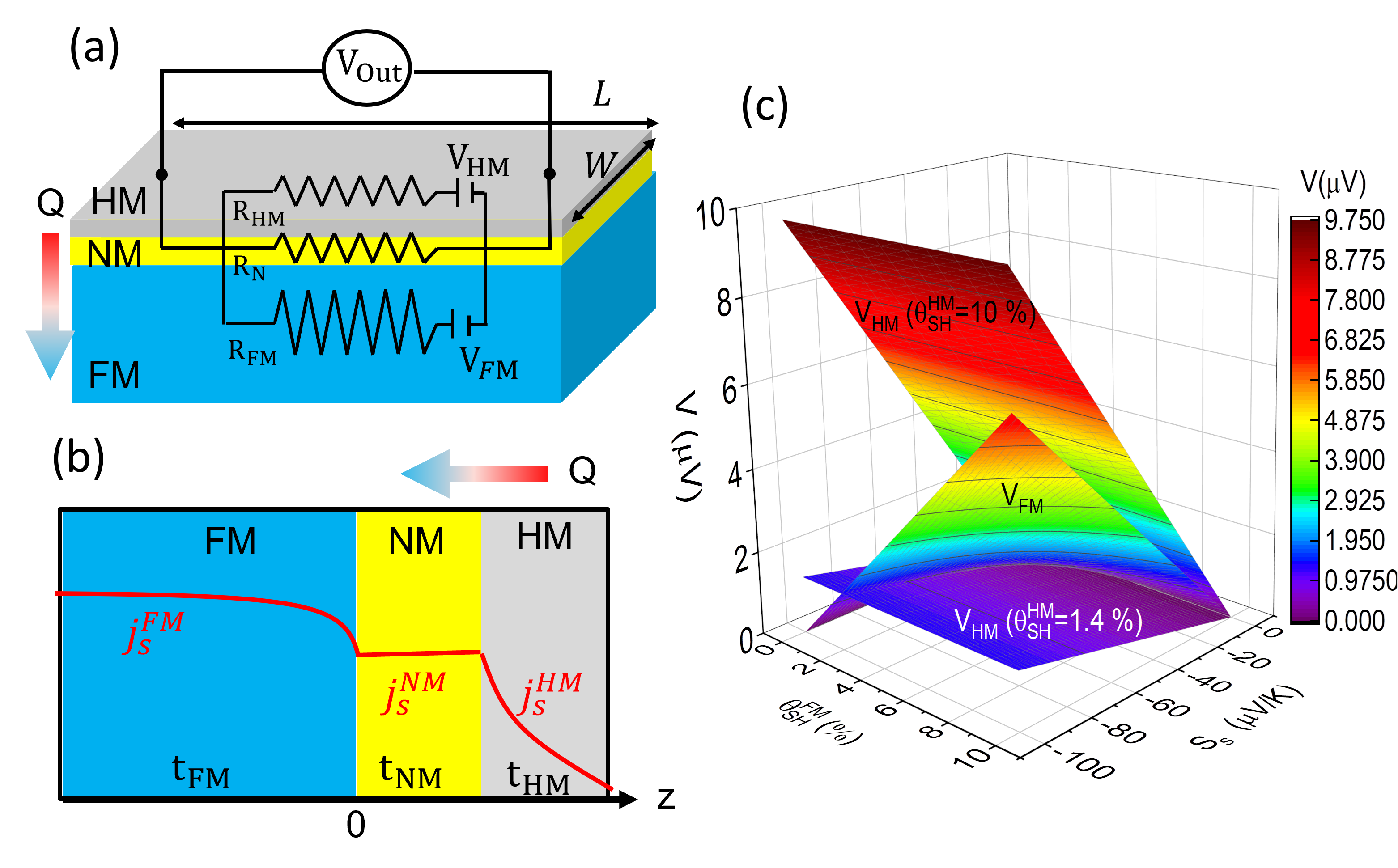}
    \caption{(a) Schematic diagram of output voltage through thermo-spin transport properties in FM/NM/HM trilayer structure. (b) Cross-section of the structure with the value of spin current density as a function of z. (c) The calculated $\rm V_{FM}$ and $\rm V_{HM}$ as a function of spin-dependent Seebeck coefficient $S_S$ and spin Hall angle $\theta_{SH}^{F}$ in the present structure for fixed spin Hall angle of HM $\theta_{HM}=1.4\%, 10\%$.
    }
    \label{2}
\end{figure*}

\section{Theory of the thermo-spin transport in FM/NM/HM}
In this section, we will discuss the thermo-spin transport properties in ferromagnet (FM)/non-magnet(NM)/heavy-magnet(HM) trilayer structures, as shown in Fig.2(a). 
Theoretically, we have expanded our study to include an in-depth analysis of the longitudinal thermo-electric effect. This phenomenon is analyzed through thermo-spin-charge conversion, a process that can be phenomenologically explained by considering the interaction of thermal spin currents with the inverse spin Hall effect in ferromagnetic and heavy metal.
When we apply the temperature gradient in z-direction for this structure, the generated spin current by the spin-dependent Seebeck effect in FM will be injected into NM and HM, and converted into a charge current in HM due to the ISHE. In addition, the ANE voltage can also be interpreted by the ISHE of thermo-spin current in FM.
The generation, injection and conversion of spin current in the FM/NM/HM structure can be described by the Valet-Fert model with consideration of the thermo-spin effect for the one-dimensional case as follows \cite{1993Valet}:
\begin{align}
\nabla^2(\sigma_\uparrow\mu_\uparrow+\sigma_\downarrow\mu_\downarrow)=0\\
    \nabla^2(\mu_\uparrow-\mu_\downarrow)=\frac{(\mu_\uparrow-\mu_\downarrow)}{\lambda^2}-e\left (\frac{dS_{\rm S}}{dT}(\nabla T)^2+S_{\rm s}\nabla^2T\right )
 \end{align}
where $\mu_{\uparrow,\downarrow}$ are the spin-dependent electrochemical potentials, and their difference $\mu_s=\mu_\uparrow-\mu_\downarrow$ is called spin accumulation. 
where, $\sigma_{\uparrow}=\sigma\frac{(1+P)}{2}$ and $\sigma_{\downarrow}=\sigma\frac{(1-P)}{2}$ are the spin-dependent conductivity. P=$\frac{\sigma_{\uparrow}-\sigma_{\downarrow}}{\sigma_{\uparrow}+\sigma_{\downarrow}}$ is the spin polarization of materials. 
 $\lambda$ is the spin-diffusion length. $S_{\uparrow,\downarrow}$ are the Seebeck coefficient for up and down spins,  respectively \cite{TULAPURKAR2010,HUEBENER1972,Duan2019,Adachi_2013}. $S_{\rm s}=S_\uparrow-S_\downarrow$ is the spin-dependent Seebeck coefficient. if we ignore the high order source terms $-e(\frac{dS_{\rm s}}{dT}(\nabla T)^2+S_{\rm s}\nabla^2T)$ in the derivation.
  The general solution of Eqs.(1)-(2) for $\mu_{\uparrow,\downarrow}$ is:
\begin{equation}
    \mu_{\uparrow,\downarrow}(z)=A+Bx\pm Ce^{\frac{z}{\lambda}}\pm D e^{\frac{-z}{\lambda}}
    \label{eq3}
\end{equation}
where the parameters A-D could be solved in different regions with the boundary conditions. 

The spin-dependent current could be expressed as following \cite{Slachter2010ThermallyDS}:
\begin{equation}
j_{\uparrow,\downarrow}=-\sigma_{\uparrow,\downarrow}\left( \frac{\nabla \mu_{\uparrow,\downarrow}}{e}+S_{\uparrow,\downarrow}\nabla T\right )
\end{equation}

Then, we solve Eq.(1)-(2) with the boundary conditions to obtain the spin current $j_{\rm s} =j_{\uparrow}-j_{\downarrow}$ in the FM, NM and HM as follows:
\onecolumngrid 
\begin{equation}
    \begin{aligned}
    j_{\rm s}^{\rm FM}=\frac{-S_{\rm s}\nabla T(1-P^2)/2\left (\lambda_{\rm FM}\lambda_{\rm HM}\sigma_{\rm FM}\sigma_{\rm NM}^2(e^{\frac{2t_{\rm NM}}{\lambda_{\rm NM}}}-1)+A^{\prime}+\lambda_{\rm FM}\lambda_{\rm NM}\sigma_{\rm FM}\sigma_{\rm NM}\sigma_{\rm HM}(e^{\frac{2t_{\rm NM}}{\lambda_{\rm NM}}}+1)-B^{\prime}\right )}{(e^{\frac{2t_{\rm NM}}{\lambda_{\rm NM}}}-1)C^{\prime}+(e^{\frac{2t_{\rm NM}}{\lambda_{\rm NM}}}+1)D^{\prime}}\\
    A^{\prime}=\lambda_{\rm NM}^2\sigma_{\rm FM}^2\sigma_{\rm HM}(P^2-1)\left (1-e^{\frac{2t_{\rm NM}}{\lambda_{\rm NM}}}-e^{\frac{z}{\lambda_{\rm FM}}}+e^{\frac{2t_{\rm NM}\lambda_{\rm FM}+\lambda_{\rm NM}z}{\lambda_{\rm FM}\lambda_{\rm NM}}}\right )\\
    B^{\prime}=\lambda_{\rm NM}\lambda_{\rm HM}\sigma_{\rm FM}^2\sigma_{\rm NM}(P^2-1)\left (1-e^{\frac{2t_{\rm NM}\lambda_{\rm FM}+\lambda_{\rm NM}z}{\lambda_{\rm FM}\lambda_{\rm NM}}}+e^{\frac{2t_{\rm NM}}{\lambda_{\rm NM}}}-e^{\frac{z}{\lambda_{\rm FM}}}\right )\\
    C^{\prime}=(1-P^2)\lambda_{\rm NM}^2\sigma_{\rm FM}\sigma_{\rm HM}+\lambda_{\rm FM}\lambda_{\rm HM}\sigma_{\rm NM}^2\\
    D^{\prime}=(1-P^2)\lambda_{\rm NM}\lambda_{\rm HM}\sigma_{\rm FM}\sigma_{\rm NM}+\lambda_{\rm FM}\lambda_{\rm NM}\sigma_{\rm NM}\sigma_{\rm HM}
    \end{aligned}
\end{equation}

\begin{small}
    \begin{align}
     j_{\rm s}^{\rm NM}=\frac{-S_{\rm s}\nabla T(1-P^2)/2\lambda_{\rm FM}\sigma_{\rm FM}\sigma_{\rm NM}e^{\frac{-t_{\rm NM}}{\lambda_{\rm NM}}}\left (e^{\frac{t_{\rm NM}+z}{\lambda_{\rm NM}}}(\lambda_{\rm NM}\sigma_{\rm HM}-\lambda_{\rm HM}\sigma_{\rm NM})+e^{\frac{3t_{\rm NM}-z}{\lambda_{\rm NM}}}(\lambda_{\rm NM}\sigma_{\rm HM}+\lambda_{\rm HM}\sigma_{\rm NM})\right )}{\left (e^{\frac{2t_{\rm NM}}{\lambda_{\rm NM}}}-1\right )\left ((1-P^2)\lambda_{\rm NM}^2\sigma_{\rm FM}\sigma_{\rm HM}+\lambda_{\rm FM}\lambda_{\rm HM}\sigma_{\rm NM}^2\right )+\left (e^{\frac{2t_{\rm NM}}{\lambda_{\rm NM}}}+1\right )((1-P^2)\lambda_{\rm NM}\lambda_{\rm HM}\sigma_{\rm FM}\sigma_{\rm NM}+\lambda_{\rm FM}\lambda_{\rm NM}\sigma_{\rm NM}\sigma_{\rm HM})}
    \end{align}
\end{small}

\begin{small}
\begin{equation}
    j_{\rm s}^{\rm HM}=\frac{-S_{\rm s}\nabla T (1-P^2)\lambda_{\rm FM}\lambda_{\rm NM}\sigma_{\rm FM}\sigma_{\rm NM}\sigma_{\rm HM}e^{\frac{t_{\rm NM}\lambda_{\rm NM}+t_{\rm NM}\lambda_{\rm HM}-\lambda_{\rm NM}z}{\lambda_{\rm NM}\lambda_{\rm HM}}}}{\left (e^{\frac{2t_{\rm NM}}{\lambda_{\rm NM}}}-1\right )((1-P^2)\lambda_{\rm NM}^2\sigma_{\rm FM}\sigma_{\rm HM}+\lambda_{\rm FM}\lambda_{\rm HM}\sigma_{\rm NM}^2)+\left(e^{\frac{2t_{\rm NM}}{\lambda_{\rm NM}}}+1\right )((1-P^2)\lambda_{\rm NM}\lambda_{\rm HM}\sigma_{\rm FM}\sigma_{\rm NM}+\lambda_{\rm FM}\lambda_{\rm NM}\sigma_{\rm NM}\sigma_{\rm HM})}
\end{equation}
  \end{small}
\twocolumngrid 
where, $\sigma_{\rm FM}$, $\sigma_{\rm NM}$ and $\sigma_{\rm HM}$ are the conductivity of FM, NM and HM, respectively.   $\lambda_{\rm FM}$, $\lambda_{\rm NM}$ and $\lambda_{\rm HM}$ are the spin diffusion length of FM, NM and HM, respectively. $t_{\rm NM}$ is the thickness of NM.

We provide a schematic of the spin current density in the trilayer system as a function of z in Fig2.(b). The flat current density value in the Cu layer means the loss of spin current due to the spin scattering in the Cu wire is negligible. This also revealed that the insertion of Cu could not reduce the spin injection efficiency into the HM layer. 
The spin current density in HM and FM is not uniform. For convenience of calculation, we use the average spin current density $J_{\rm s}^{\rm HM}=\frac{1}{t_{\rm HM}}\int_{t_{\rm NM}}^{t_{\rm NM}+t_{\rm HM}}j_{\rm s}^{\rm HM}d t_{\rm HM}$ in heavy metal, $J_{\rm s}^{\rm FM}=\frac{1}{t_{\rm FM}}\int_{-t_{\rm FM}}^0j_{\rm s}^Fdt$ in ferromagnet, where $t_{\rm FM}$, $t_{\rm HM}$ is the thickness of the FM and HM.\par
The voltage generated in FM and HM can be calculated by $J_{\rm s}^{\rm FM}$ and $J_{\rm s}^{\rm HM}$ as follows:
\begin{align}
    V_{\rm FM}=J_{\rm s}^{\rm FM}\theta_{\rm SH}^{\rm FM}\rho_{\rm FM}L\\
    V_{\rm HM}=J_{\rm s}^{\rm HM}\theta_{\rm SH}^{\rm HM}\rho_{\rm HM}L
\end{align}
where $L$ is the length of HM and FM, $\theta_{\rm SH}$ is the spin hall angle, which is the conversion efficiency between the spin current and charge current. $\rm \rho_{\rm FM},\rho_{\rm HM}$ are the resistivities of FM and HM, respectively.
It's clear to see that the $\rm V_{\rm FM}$ and $\rm V_{\rm HM}$ are mainly dominated by the thermo-spin current and spin Hall angle. 
To understand the thermo-voltage contribution in the FM and HM layer, we also calculate the $V_{\rm FM}$ and $V_{\rm HM}$ as a function of spin-dependent Seebeck coefficient $S_S$ and spin Hall angle $\theta_{\rm SH}^{\rm FM}$ in the present structure for a fixed spin Hall angle of HM ($\theta_{\rm HM}$), as shown in Fig. 2(c). The related parameters are given in Table \ref{tab:my_label}. We can clearly see that there is a crossing between the $V_{\rm FM}$ and the $V_{\rm HM}$ with the lower spin Hall angle 1.4\% (the lower estimated value of Pt). This result indicates that the lower $S_{\rm s}$  or $\theta_{\rm FM}$ will induce lower ANE voltage, which could be much lower than the inversed Hall voltage in the HM layer. However, if the much higher spin Hall angle 10\% (the higher estimated value of Pt), all the values of $V_{\rm HM}$ are much larger than that of $V_{\rm FM}$. 
In designing transverse thermo-electric devices, careful consideration of relevant parameters is essential when selecting materials.

\begin{table}
    \centering
    \caption{Parameters for the calculation of $V_{\rm HM}$ and $V_{\rm FM}$}
    \begin{tabular}{l|l}
    \hline
    \textbf{Parameters} & \textbf{Values} \\
    \hline
    \\
        $\rho_{\rm FM}$ (CoFeAl) & $4.5\times 10^{-7}\ \Omega\cdot m$\ \cite{Hu2014}\\
        $\rho_{\rm NM}$ (Cu) & $2.95\times 10^{-8}\ \Omega\cdot m$\ \cite{cui2022enhanced}\\
        $\rho_{\rm HM}$ (Pt) & $1.56\times 10^{-7}\ \Omega\cdot m$\ \cite{kimura2007room}\\
        $t_{\rm FM}$ & $35\ nm$\\
        $t_{\rm NM}$ & $5.4\ nm$\\
        $t_{\rm HM}$ & $4.4\ nm$\\
        $\theta_{\rm SH}^{\rm HM}$ (Pt) &$1.4\%$\ \cite{PhysRevLett.111.066602}, $10\%$\ \cite{PhysRevB.82.214403} \\
        $\nabla T$ & $10^{5}\ K/m$ \\
        $W$ & $20\ \mu m$ \\
        $L$ & $225\ \mu m$\\
        $P$ (CoFeAl) & 0.62\ \cite{Hu2014}\\
        $\lambda_{\rm FM}$ (Py, CoFeAl)& $2\ nm$\ \cite{cui2022enhanced}\\
        $\lambda_{\rm NM}$ (Cu) & $450\ nm$\ \cite{Hu2014}\\
        $\lambda_{\rm HM}$ (Pt) & $5\ nm$\ \cite{lamdbdaPt}\\
    \hline
    \end{tabular}
    \label{tab:my_label}
\end{table}

\begin{figure*}[htp]
\centering  
\includegraphics[width=6.3in]{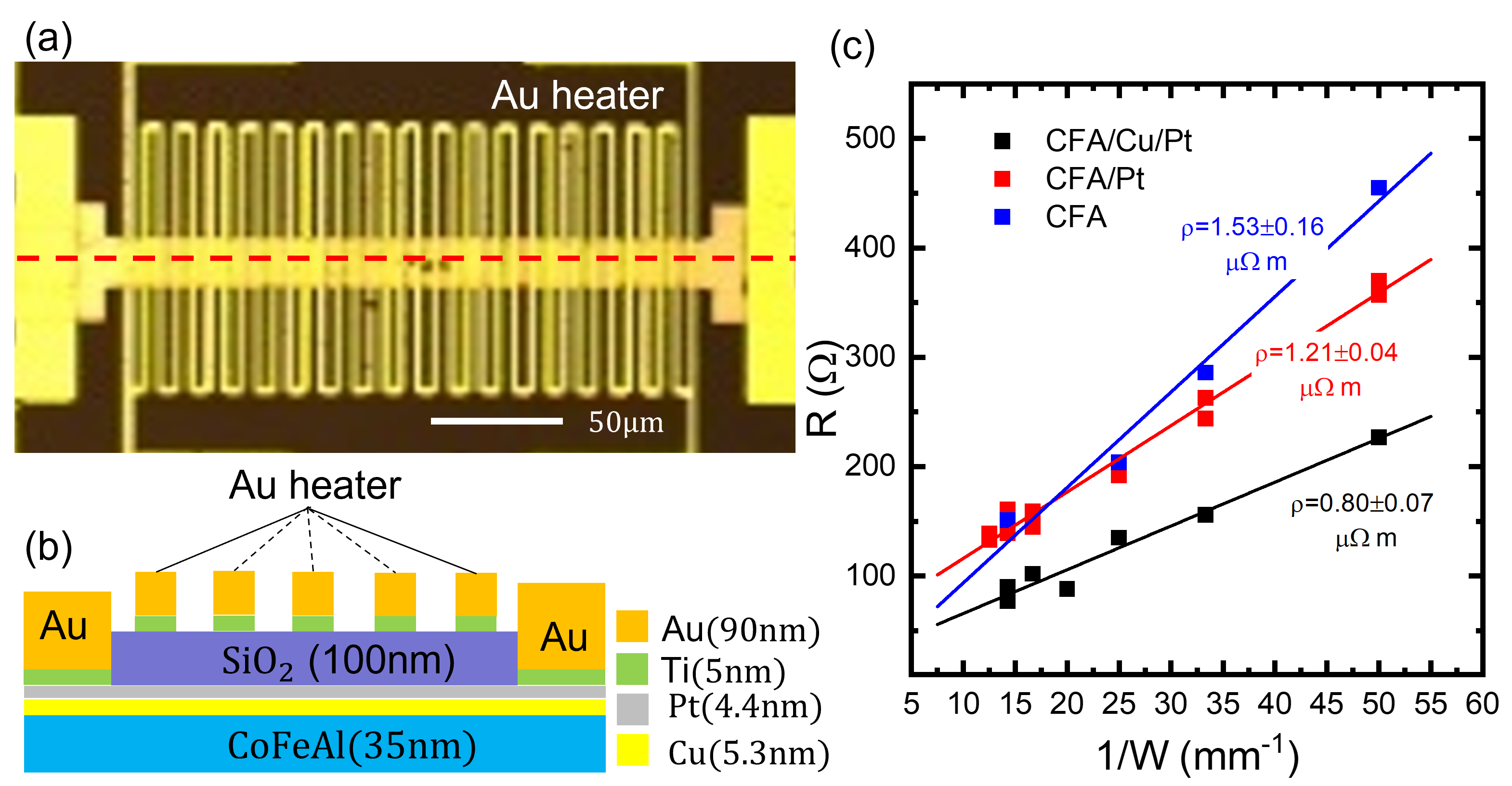}
\caption{(a) The top view picture of on thermo-spin device with 225$\mu m$ in length and 20 $\mu m$ in width. The zigzag Au wire is the heater providing a temperature gradient in the top layer. (b) Side view diagram of the device. For the device, the thickness of CoFeAl, Cu and Pt are 35 $nm$, 5.3 $nm$ and 4.4 $nm$, respectively. The Au on the left and right sides is used as the detection electrode.  (c) The resistances as a function of the device's $1/W$ ($W$ is the width of the devices) for various kinds of structures. Their resistivities could be obtained by the linear fitting of $1/W$. 
}
\label{3}
\end{figure*}

Then, the output voltage $V_{\rm out}$, which is coming from the two voltage sources $V_{\rm HM}$ and $V_{\rm FM}$, is given as follows by considering the inner short circuit current :
\begin{equation}
V_{\rm out}=\frac{\rho_{\rm HM}\rho_{\rm NM} t_{\rm FM} V_{\rm FM}+\rho_{\rm FM}\rho_{\rm NM} t_{\rm NM} V_{\rm HM}}{\rho_{\rm HM}\rho_{\rm NM} t_{\rm FM}+\rho_{\rm FM}\rho_{\rm NM} t_{\rm HM}++\rho_{\rm FM}\rho_{\rm HM}t_{\rm NM}}
\end{equation} 
It's clear to see that the output voltage is free of the width of the trilayer. But it shows the linear function with the length L of the structure. 

The output power of the device is given as follows:
\begin{equation}
   P_{\rm out}=(\frac{V_{\rm out}}{R_{\rm out}+r})^2\times R_{\rm out}
\end{equation} where $r=\rho_{\rm Tri}\ L/(d_{\rm Tri}\ W)$ the internal resistance and $R_{\rm out}$ is the external resistance. The internal resistance of the three-layer structure could be significantly modified by $W$ for a fixed L. However, $V_{\rm out}$ does not change with the width $W$.  This is an advantage for optimising the devices for the fixed output voltage with the maximum output power by adjusting the internal resistance. To confirm such properties, we will fabricate the thermo-spin conversion devices with CoFeAl/Cu/Pt structure.

\section{Experimental results}
Here, we fabricated a batch of devices with different widths of trilayer by using the optical lithography with the mask aligner function and a wet lift-off method. 
First, the CoFeAl/Cu/Pt trilayer is deposited in the ultra-high vacuum electron beam evaporator without breaking the vacuum. 
The CoFeAl is about 35 nm, which is much thicker than its spin diffusion length. The Cu is about 5.3 nm, which is much smaller than its spin diffusion length even at room temperature. The Pt layer is about 4.4 nm, which is almost the same as its spin diffusion length.  After that, 100 nm SiO$_2$ is grown by thermo-atomic layer deposition method on the top of the trilayer film to isolate the heater. Finally, the 90 nm Au film was fabricated by e-beam evaporation for the electrode probes and zigzag shape heater. To enhance the adhesion of the Au wire, 5 nm Ti was deposited before Au deposition. One final fabricated device is shown in Fig.3(a).
\begin{figure*}[htp]
\centering  
\includegraphics[width=6.3in]{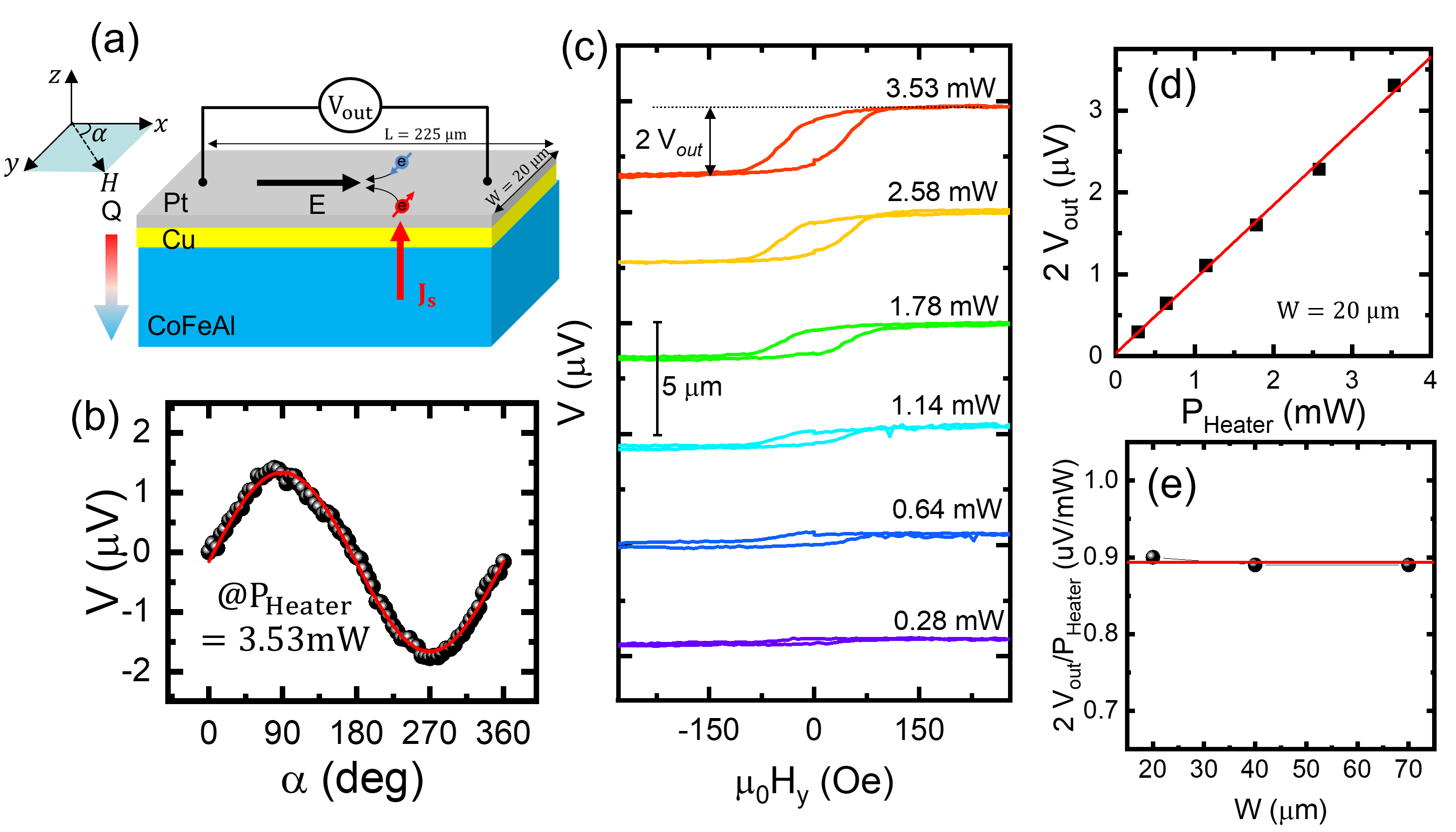}
\caption{(a) The schematic diagram of thermoelectric device model. The heat current is applied in the longitudinal direction to generate longitudinal thermo-spin current in CoFeAl. The thermo-spin current could be injected into Cu and Pt and converted to charge current in CoFeAl and Pt, resulting in charge accumulation and generating a detectable electric potential $V_{\rm out}$. 
(b) The magnetic field angle-dependent transverse thermo-voltage diagram. (c) Variation of $V_{\rm out}$ with magnetic field intensity at different heating power. (d) When W=20 $\mu m$, 2$V_{\rm out}$ changes with heating power ($P_{\rm heater}$).(e) The relationship between $V_{\rm out}/P_{\rm heater}$ and width W.}
\label{4}
\end{figure*}
To obtain the resistivities of various types of structures, we evaluated the 1/W dependence of the resistances for the CoFeAl, CoFeAl/Pt and CoFeAl/Cu/Pt sheets, as shown in Fig.3(c). The resistances are measured using the two-terminal method. 
It can be seen that the resistance of the device with three structures varies linearly with 1/W. By linear fitting the data, the obtained resistivities $\rho_{\rm CFA}=1.53\pm0.16\ \mu\Omega m$, $\rho_{\rm CFA/Pt}=1.21\pm0.04\ \mu\Omega m$, $\rho_{\rm CFA/Cu/Pt}=0.80\pm0.07\ \mu\Omega m$,  respectively.
Obviously, inserting the Cu between CoFeAl and Pt layers can significantly reduce the resistivity of the structure. 

Then, we evaluated the thermo-spin-charge conversion effect of the CoFeAl/Cu/Pt devices with the illustration shown in Fig.4(a). The heat current flows from the top to down, and the thermal spin could be injected from the CoFeAl layer to the Cu and diffuse to the Pt layer. And the voltage will be generated in the $x$ direction due to the inverse spin Hall effect in CoFeAl and Pt. The detected voltage is related to the spin vector of the spin current, which is dominated by the magnetic momentum of CoFeAl. To confirm such an effect, we measured the external magnetic field angle dependence of the transverse thermo-voltage at a fixed heating power of 3.53 mW, as shown in Fig.4(b). $\alpha$ is the angle between the external magnetic field H with the $x$-axis. The out-of-voltage could be well-fitted by the sine function. 
We also plot the V-H curves for various heating power with the sweeping direction of the magnetic field paralleling $y$-axis in Fig.4(c). Here, the transverse thermo-voltage ($V_{\rm out}$) is defined as the $[V(\rm \mu_0 H +)$$-$$V(\rm \mu_0 H -)]/2$. It's clear to see the transverse thermo-voltage is significantly enhanced by increasing the heating power. Then, we plot $2 V_{out}$ as a function of heating power ($P_{\rm heater}$) for the device with $W$ = 20 $\mu m$ in Fig.4(d). The $2 V_{\rm out}$ is well-fitted using the linear relation with the heating power. To fairly compare the device's properties, we also obtained the value of $2V_{\rm out}/P_{\rm heater}$ for various devices with different widths W in Fig.4(e).  Using a linear fitting with a fixed zero slope, we determined the fixed intercept to be 0.98, with an error margin of less than 4\%. Hence, we can conclude that the ratio of $\rm 2V_{\rm out}/P_{\rm heater}$ remains constant. It does not depend on the width of the trilayer. This is also consistent with our theoretical analysis. 

\section{Conclusion}
We conducted a theoretical analysis of thermo-spin generation, transportation and spin-charge conversion in vertical FM/NM/HM structure. Our findings revealed that the output transverse thermo-voltage is not influenced by the structure's width, but varies with the length of the structure in a linear manner. The results of thermo-spin devices with a CoFeAl/Cu/Pt structure verify our predictions. This demonstrates the potential to design flexible thermo-spin conversion devices using FM/NM/HM structures.

\begin{acknowledgments}
This work is partially supported by National JSPS Program for Grant-in-Aid for Scientific Research (S)(21H05021), and Challenging Exploratory Research (17H06227) and JST CREST (JPMJCR18J1).
\end{acknowledgments}

\bibliographystyle{unsrt}
\bibliography{Bib}

\end{document}